\documentclass[12pt]{article}
\usepackage{amsmath}
\usepackage{amsfonts}
\def\ie{{\em i.e.\,}}

\begin{document}
\title{Genetic information and quantum gas}
\author{M.V.Altaisky\thanks{Joint Institute for Nuclear Research,
 Dubna, 141980, Russia;
 and Space Research Institute RAS, Profsoyuznaya 84/32, 
Moscow, 117810, Russia, altaisky@mx.iki.rssi.ru}
\and F.P.Filatov\thanks{ Research Center for Toxicology and Hygienic 
Regulation of Biopreparations of the Ministry of Public Health of the Russian
     Federation, 102a Lenin Str., Serpukhov, 142283, Russia; and 
National Hematology Research Center of  RAMS,4a Novo-Zykovsky pr., 
Moscow, 125167, Russia, ffelix@blood.ru}}
\date{}
\maketitle
\begin{abstract}
Possible explanation of the 64/20 redundancy of the triplet genetic code 
based on the assumption of quantum nature of genetic information is proposed. 
\end{abstract}

  The redundancy of the genetic code, where 64 possible nucleotide 
triplets code only 20 amino acids used as building blocks for all proteins,
 is well known. For a non-degenerate triplet  code written in a four letter 
nucleotide alphabet ({\sl ATCG}), the naive expectation of the possible coding 
repertoire is equal to the number of all sequences of $r$  
elements  ($r=3$, triplet) with each element taken from a set containing
$k$ elements ($k=4$, number of bases):
\begin{equation}
A^k_r = k^r, \quad{or\ } A^4_3 = 64.
\label{classic}
\end{equation}

In reality, only 20 amino acids are encoded, thus the code 
is degenerate because different triplets code the same amino acid. In each 
triplet the first two letters are more significant than the third one, 
but without full degeneracy: otherwise we would have only $4^2=16$ amino 
acids instead of 20. The endeavor of answering the question on the origin
 of $64/20$ degeneracy utmost inevitably leads to the conclusion that on the 
early 
stages of evolution codon was functioning as a set of nucleotides, rather than 
an ordered sequence. If it is the case, the total number of functionally 
different combinations (functional keys) should be equal to the number of  
sets equivalent with respect to permutations of $r$ elements with each element taken from a $k$-element set:
\begin{equation}
C^k_r = \frac{(k+r-1)!}{r!(k-1)!}, \quad{or\ } 
C^4_3 = \frac{(4+3-1)!}{3!(4-1)!}=20.
\label{quantum}
\end{equation}

The coincidence between combinatoric rule \eqref{quantum} and the 
redundancy $64/20$ was noted long ago by Gamov \cite{gamov}, who 
stressed it as a question to be answered in the future. Recently Patel 
\cite{patel}, in relation to quantum computations, suggested that the 
choice of 20 amino acids is a result of optimization with respect to 
a quantum search algorithm, like that proposed by Grover\cite{grover}.
The Grover's algorithm relates the number of objects $N$ that can be 
distinguished by a number of yes/no queries, $Q$, according to 
\begin{equation}
(2Q+1) \sin^{-1}(1/\sqrt N) = \pi/2.
\label{nq}
\end{equation}
For $Q=3$ the exact solution is $N=20.2$. 
Thus, from the standpoint of quantum information  theory, $N=20$, 
being very close to $20.2$, provides 
a quantum database search with minimal possible errors asking 3 questions 
only. 

We would like to  propose another interpretation of the genetic code 
redundancy 64/20. The formula \eqref{quantum} exactly coincides with the 
number of quantum states of a Bose gas of $r=3$ particles  with $k=4$ 
possible quantum states, \ie with the number 
of arrangements of $r$ indistinguishable particles into $k$ states.  

The concept of an 
ideal Bose gas is an  idealization used for  
adiabatically isolated systems of identical quantum particles.
  To some extent, we can suggest that this  idealization works for the 
transcription and translation of genetic information.  
Taking into account that the typical 
codon size is about 0.5 nm, from the Heisenberg uncertainty principle 
\begin{equation}
\Delta x \Delta p \approx \hbar,
\end{equation} we can easily infer that 
for the typical size of a nucleotide 
$$\Delta p \sim \frac{1.05\cdot 10^{-27}erg\cdot sec}{1.7\cdot 10^{-8}cm} 
\approx 6.2\cdot10^{-20}\frac{g\cdot cm}{sec}.$$ 
For the typical mass of a hydrogen atom the energy associated with 
$\Delta p$ is 
\begin{equation}
\Delta E = \frac{(\Delta p)^2}{2m} \approx 
\frac{(6.2\cdot10^{-20}\frac{g\cdot cm}{sec})^2}{2\cdot 1.67\cdot 10^{-24}g} 
             \approx 1.2\cdot 10^{-15} erg 
\label{de}
\end{equation}
For comparison, the typical energy of the hydrogen bond (in water) 
is $1\frac{kkal}{mole} \approx 7\cdot 10^{-14}erg$ \cite{cell}.
Thus, the difference between the quantum fluctuations energy and the typical 
energy of hydrogen bond is one order only, and this small contribution 
might be significant for the formation of fine molecular structures. 
Due to the quantum fluctuation contribution, the processing by any biological 
agent of a single nucleotide by means of formation of hydrogen 
bonds is less reliable than the simultaneous processing of a nucleotide 
complex. 
For a codon, the size of which 
is roughly 3 times larger than that of a nucleotide, the quantum fluctuation energy  is one 
order less, and that provides a reliable transmission of information.     
In other words, we can say that reading/writing of genetic information 
occurs by means quantum transitions of adiabatically isolated 
codon-anticodon pairs, i.e. in the quantum system of 6 nucleotides, three 
of those carry information content. 

For complete analogy with quantum Bose gas, one could expect the 
codons with same nucleotide content (but different order of nucleotides)   
to code the 
same nucleic acid. The possible explanation for this does not happen is, 
that at early stages of the evolution this was really so, 
and each triplet  was functioning as just a set  
of proto-nucleotides without any internal ordering, but later in course 
of the evolution with changing structure of nucleotides 
the symmetry 
with respect to permutation was spontaneously broken: the system 
evolved to a new ground state with less energy. What we observe now might 
be a result of this  symmetry breaking. If so, this  
symmetry breaking should be non-degenerate: it maps 20 quantum states 
before the breaking to 20 states after it. It is important 
that if the proposed concept of  adiabatically isolated quantum gas of 
nucleotides is valid, it guarantees exactly 20 possible states: not less,  
not more.        
\vskip\baselineskip
\centerline{Aknowledgement}
The authors are thankful to  Dr. D.~Chakalov for some bibliographical 
references related to this subject. 
  
This work was supported in part by ISTC project  1813p/2001.

\end{document}